# Review of Replication Schemes for Unstructured P2P Networks

**Sabu M. Thampi†, Chandra Sekaran K††**

†L.B.S Institute of Technology for Women, Kerala-695012, India
††National Institute of Technology Karnataka, Surathkal, Karnataka-575025, India
*smtlbs@yahoo.co.in, kch@nitk.ac.in*

*Abstract*- **To improve unstructured P2P system performance, one wants to minimize the number of peers that have to be probed for the shortening of the search time. A solution to the problem is to employ a replication scheme, which provides high hit rate for target files. Replication can also provide load balancing and reduce access latency if the file is accessed by a large population of users. This paper briefly describes various replication schemes that have appeared in the literature and also focuses on a novel replication technique called Q-replication to increase availability of objects in unstructured P2P networks. The Q-replication technique replicates objects autonomously to suitable sites based on object popularity and site selection logic by extensively employing Q-learning concept.**

## I. Introduction

P2P traffic keeps on increasing and its share of entire network traffic is escalating quickly. The major operations associated with decentralized unstructured P2P network can be summarized into two phases: (i) query phase and (ii) download phase. In query phase, several query packets pass through the network searching for the target objects. The heterogeneity of these query packets creates a local traffic disparity and congestion. The downloading of large objects in the download phase in response to requests also causes congestion in nodes. One proficient method for forestalling this load concentration is replication of the target objects into various sites. Replication increases object availability and fault tolerance. Single node failures, like crashes of nodes, can be tolerated as faults within the system as a whole facilitated with the help of the redundancy introduced by replicas. If a host of a replica fails, requestors may access another host with a replica. Data replicated at more than one site facilitate to minimize the number of hops before the data are found.

Replicating objects to multiple sites has several issues such as selection of objects for replication, the granularity of replicas, and choosing appropriate site for hosting new replica [1]. The existing replication techniques address these issues differently. Excessive replication can cause wastage of network and peer resources and at the same time, scarcity of resources decreases the search success rate and increases the search delay. Two important aspects of replication—selection of file for replication and selection of site for hosting new replica—have a direct impact on the performance of the system. Suitable criteria should be followed for selecting a file for replication. If popular files are not replicated appropriately, overwhelming requests from peers can cause network congestions and slow download speed. Based on the location selection logic for hosting new replica, replicated copies should be placed in proximity to peers who are likely to request the resource. This allows peers to be able to search and find desired resources, and reduces delays taking place during search and downloading. The replication strategy should use different characteristics of peers such as available storage and their surrounding usage environment attributes such as network bandwidth to determine which peers should be selected to perform replications and where the resulting replicas should be stored.

Majority of the existing replication methods only replicate objects to intermediate nodes between query node and target node. These replication schemes depend completely on the search path. Due to this, objects are unnecessarily replicated to low performing nodes on the search path. It is essential that the objects should not be replicated to low performing nodes since these nodes are not queried frequently by other nodes; excluding such nodes from replicating files can save bandwidth. In a network, many peers might have decided to replicate the same file at the same time. This should be managed; otherwise, the same file could be copied into nodes repeatedly. A replication scheme should be well designed to manage the frequent failure of nodes in the network to provide good success rate by maintaining replicas in other suitable peers. The various issues in replication demands a more assertive replication approach for unstructured P2P networks. This paper briefly describes various replication schemes that have appeared in the literature and also focuses on a novel replication technique called *Q-replication* to address the above issues for effectively increasing the availability of objects in the network. The Q-replication technique considers the replica selection problem (which data to replicate) and the replica placement problem (where to place them), and provides simple solutions to them. The replica selection problem deals with a suitable criterion for selecting an object from the shared storage space of a node for replication. The replica placement problem addresses the process of choosing an appropriate node for hosting a replica. As the scheme employs Q-learning for the selection of target peers; the probabilistic or random selection of target peers for hosting the replica are avoided.

The remainder of this paper is organized as follows.

Section II reviews the existing replication techniques. An overview of the Q-replication technique is given in section III. Section IV concludes the paper.

## II. Replication Techniques

*Uniform, Proportional and Square root replication* [1, 2, 3]: In the uniform strategy, replications are uniformly distributed throughout the network. For each data object approximately the same number of replicas are created. While this controls the overhead of replication, replicas may be found in places where peers do not access the files. In the proportional replication, the number of copies for each object is proportional to its query distribution. The higher the query rate of an object, the higher is the number of copies for that object. With proportional replication, on the other hand, although queries on popular data are processed efficiently, unpopular data search may take a long time degrading the overall system performance. In the case of square-root (SR) replication the number of replicas of a file 'i' is proportional to the square root of query distribution, $q_i$. Optimal replication is attained when the number of replicas per item is proportional to the square root of their popularity. Uniform and Proportional strategies have been shown to have same search space.

*Owner replication, path replication and random replication:* All the three schemes [3, 4, 5, 6] based on site selection policy replicate the found object when a query is successful. The owner replication replicates an object only at the requesting node. The number of replicas increase in proportion to the number of requests for the service. Random replication distributes the replicas in a random order rather than following the topological order. If we use random forwarding n-walkers random walk, random replication is the most effective approach for achieving both smaller search delays and smaller deviations in searches. On the other hand, to do random replication, the peer must know the information of all the peers in the logical network. This is very difficult to implement since a peer only contains information about its neighbouring peers. The path replication creates copies of an object on all nodes on the path from the providing node to the requesting node and its implementation is less complex than the random replication. It has been shown in [3] that factor of improvement in path replication is close to 3 and in random replication the improvement factor is approximately 4.

*Pull-Then-Push (PtP) replication:* With PtP replication [5], after a successful search, the requesting node enters a replicate-push phase where it transmits copies of the item to its neighbours in order to obtain square root replication. Updating the replicas can be significantly improved through an update-push phase where the node that created the copies propagates any updates it has received using similar parameters as in replicate-push. The problem with SR replication is that it requires knowledge of the query rate for each item. To improve this, in PtP, after each successful search, the item is copied to a number of nodes equal to the number of probes. The creation of replicas is delegated to the inquiring node, not the providing node. The scheme consists of two phases. The pull phase refers to searching for a data item. After a successful search, the inquiring node enters a push phase, whereby it transmits the copies of the item to its neighbours in order to force creation of replicas. In order to reach SR replication, number of replicas equal to the number of probed nodes are created. The same algorithm is used for both the push and the pull phases, so that the push phase visits approximately the same nodes the pull phase visited. Since low replicas are placed on all the nodes probed, low performing nodes may also unnecessarily receive replicas.

*Path random replication and Path adaptive replication:* Path Replication places replicas in all the peers on the path the requested data goes to the requesting peer. The number of replicas created can become very large, which eventually may be more than necessary to achieve the required search performance. Thus, some amount of the processing capability and storage capacity of the peers may be wasted, particularly on the few peers with a high degree. Path replication method coupled with a replication ratio is referred to as the path random replication method [7]. Each intermediate peer randomly determines whether or not the replica is created and placed there, based on the probability of the pre-determined replication ratio. Path adaptive replication [7] is an alternative to path random replication that adaptively determines whether or not to create a replica depending on its storage capacity. Path adaptive replication determines the probability of the replication in each peer according to the predetermined replication ratio and its resource status. Path random replication outperforms path adaptive replication in the average number of search hops. The feedback information after replication is not collected and utilised for determining the node for hosting the replica.

*Adaptive Probabilistic Replication (APRE):* APRE [8] is a distributed protocol that automatically fine-tunes the replication ratio of each shared object according to the current demand for it. APRE offers a direct response to workload changes; by generating server points in needy areas or releasing redundant replicas in areas of low demand. APRE, which combines replication with the search protocol, is based on two basic operations: Expansion and Contraction. APRE couples lookup indices together with an aging technique to identify query concentrated areas within the P2P overlay. The indices maintained for the look up process are probabilistic. On the basis of local workload computation, peers independently decide on the time and extent of replication. APRE is robust in eliminating server overloads while curtailing the communication overhead and balancing the load.

*Optimal content replication*: Optimal content replication [9] is an adaptive, fully distributed technique that dynamically replicates content in a near-optimal manner. The optimal object replication includes a logarithmic assignment rule, which provides a closed-form optimal solution to the continuous approximation of the problem.

Two algorithms are proposed: a Top-K LRU algorithm, and a Top-K Most Frequently Requested (MFR) algorithm. Top-K LRU algorithm replicates content on the fly without any a priori knowledge of object request patterns or nodal up probabilities. The idea behind the algorithm is as follows. Each object j has attractor nodes determined by the underlying P2P substrate. The object j tends to get replicated in its attractor nodes, which go up and down over time. Queries for objects also tend to get sent to attractor nodes. Thus, a query for a particular object j tends to get directed to up nodes that likely have the object. Objects get replicated on-the-fly when none of the top-K peers have the requested object. LRU lets unpopular objects remain in peers. When an unpopular object is requested, the object gets stored in one of the peers and remains there until it is ejected with LRU. If the object is very unpopular, it will likely not receive any requests during its halt in the peer, and hence waste storage space. The MFR algorithm is an alternative to Top-K LRU algorithm, which manages the storage effectively. MFR algorithm follows its own retrieval and replacement policy and makes high hit rates that are very close to optimal.

*Adaptive replication method based on peer behaviour pattern*: This method [10] uses the relevancy and usefulness of peers to determine how many replications should be made, and where to locate these replications. When a new document is registered at a peer, the peer replicates the document. The number of replications to be made depends on the relative usefulness of the neighbours of a given peer. The target peers are determined by a peer group and peer selection criterion, where a peer group is a candidate set of peers from which the target peers are selected. Four types of peer grouping are employed: placement on neighbours (PN), placement on inverted references (PIR); placement on relevant peers (PRP) and random placement. In PN method, the peer group is defined as the set of neighbours of the source peer. The neighbour peers are ranked in ascending order of their usefulness. In PIR method, a peer group is defined as the set of peers that access the source peer to obtain documents. PRP method defines a peer group as the set of peers whose queries are sent to the source peer. Random Placement defines the peer group as all peers joining the system. The target peer is chosen randomly from all peers. Using the PN method, query processing is effective when the network does not have relatively many peers. However, the performance degrades as the number of peers increase. The PN method experiences trouble with scalability. In the case of the random method, a favourite document is placed regardless of the distance between a query peer and the peer holding the document. However, the number of query results decreases as the number of peers increases. Neither the PIR method nor the PRP method is affected by an increasing number of peers, compared with the PN and random methods. For query results, the PRP technique shows superior performance than other techniques. Even though a node has several attributes, in this scheme the usefulness of a node is defined only using a single parameter. This makes the peer selection process for replication an imprecise one.

*Decentralized replication algorithms:* The decentralized replication algorithms [11] deal with storage allocation and replica placement. The process of storage allocation decides how many replicas can be produced for each file upon the limitation of storage space, and replica placement procedure decides the set of peers that are going to store those replicas of each file to achieve a reasonable level of file availability. To provide sufficient file availability, three heuristic algorithms—a random algorithm, a group partition algorithm that relies on peers' forming groups and a greedy search algorithm based on an estimated system-level file availability target—are presented. The three replication schemes employ the erasure-coded blocks for replication. The random algorithm does not require any knowledge of peer availability, and gives each file the same stretch factor and equal opportunity in selecting peers. The group partition algorithm can achieve lower variance in file availability, hence may be a good choice if fairness of file availability is important. The greedy algorithm can achieve higher availability especially when peers share a small amount of storage space for replication and when high available peers in the system are rare. The success of the algorithms depends on the failure rate of peers in the network.

*Autonomous replication using erasure codes*: Autonomous replication using erasure codes [12] uses randomized decisions extensively together with the application of erasure codes to tolerate autonomous peer actions. Each member of a community hoards some subset of the shared files entirely on their local storage, called the member's *hoard set*, and pushes replicas of its hoard set to peers with excess storage using an erasure code. The basic steps of the algorithm are: (i) each member advertises the unique IDs of the files in its hoard set and the fragments in its replication store in the global index. Each member also advertises its average availability in the global directory, (ii) each member periodically estimates the availability of its hoarded files and the fragments in its replication store, (iii) periodically, say every $T_r$ time units, each member randomly selects a file from its hoard set; the member does this by generating a random erasure coded fragment of the file and pushes it to a randomly chosen target, and (iv) the target accepts and saves the incoming fragment if there is sufficient free space in its replication store. If there is insufficient space, it either rejects the replication request or ejects enough fragments to accept the new fragment. Victims are chosen using a weighted random selection process, where more highly available fragments are more likely to be chosen. This method minimizes the bandwidth costs in accessing the files provided that the availability of nodes should be high. The amount of replication of each file is proportional to the frequency of access to that file. The presence of a small number of highly available members can significantly reduce the replication necessary to achieve practical availability levels.

*Dynamic replication schemes:* Dynamic replication [13], which is used in superpeer P2P architecture, takes the cost of searching a data item and successfully replicates the most frequently accessed data files based on the access probabilities. Two novel techniques to share the load using replication techniques are proposed: periodic push-based replication (PPR) and on-demand replication (ODR). In PPR, the hosting super-peer periodically sends replicas of the most frequently accessed files to remote super-peers on the basis of global access frequency. By replicating, the hop count to search for a file is reduced. A super-peer receiving a replica also informs its neighbouring super-peers about the replica through a restricted gossiping algorithm. ODR performs replication based on local access frequencies. This technique allows super-peers to dynamically adapt to changes in access behaviour, however, it is greedy as each super-peer tries to perform replication based on its own needs rather than replicating from a global perspective as done in PPR.

*Dynamic model-driven replication:* This is a decentralized model, which is used for creating replicas dynamically in an unreliable P2P system [14]. The model is able to predict the required number of replicas in the system with reasonable accuracy. Each peer in the system possesses a model of the P2P storage system that it can use to determine how many replicas of any file are needed to maintain desired availability. Each peer applies this model to information it has about system state and replication status of its files to determine if, when, and where new replicas should be created. The system works as follows. Each node in the network is authorized to create replicas for the files it stores. Although the technique is probability based, it is effective in predicting the required number of replicas in the system. This approach relies on a resource discovery service to find various parameters.

*Index replication:* The objective of index replication technique [15] is to improve the search effectiveness for rare items, and reduce the bandwidth overhead incurred in superpeer based P2P networks. It explores the use of multi-hop index replication, which can significantly improve the effective search space. In two-hop index replication scheme, each node sends its index to all of its one-hop neighbours in its routing table. All of the one-hop neighbours, in turn, forward this index to all of their one-hop neighbours except the source node. This strategy effectively reduces to a two-hop flooding of indices around the nodes. Two variants of two-hop index replication are used: *SR replication* and *constant replication*. In *SR replication*, each node performs one-hop replication. Supernodes then replicate the indices of their one-hop neighbours to a random subset of their supernode neighbours. This is simple one-hop replication augmented with SR replication only at the supernodes. The advantage is reduction in the amount of replication and its cost. Finally, in *constant replication* each supernode does two-hop index replication to a constant number of supernode neighbours. After each node does one hop index replication, supernodes propagate the index to only a constant number of their supernodes. This reduces indexing load on supernodes. The overhead incurred due to increase in messages is higher than that of one-hop replication techniques.

## III. Q-Replication

The Q-replication scheme is distributed, and employed without the coordination of centralized servers. The technique employs Q-learning in various stages of replication. The process of replication involves selection of suitable objects based on popularity and selection of target nodes for hosting the replica. Based on the parameter values appropriate variables are modified.

*A. Selection of objects for replication*

The objects are chosen for replication according to their popularity. The frequently accessed objects from the shared storage space of a node are treated as popular objects. These files are ranked according to their popularity. The details are stored in a table, which contains the object name along with its rank and the status of replication. The value of the rank represents the popularity of the object. The high value rank denotes a most popular item. The status field facilitates to identify already replicated files in a node. The system regularly (e.g. for every 50 requests received by a node) updates the popularities of all the objects in the nodes based on the incoming queries. The popularity update process $P_f(t+1)$ at a time 't' relies on the number of requests received for the object $R_q(t)$ and the total number of requests received by the node $N_q(t)$ and the present value of popularity $P_f(t)$. The popularity is modified as

$$P_f(t+1) = P_f(t) + \eta \left[ \left( \frac{R_q(t)}{N_q(t)} \right) * 100 \right],$$

where the value of $P_f \geq 0$ and the value of constant $\eta$ is in between zero and one. The values thus modified are written into a table (*popularity table*) after removing the existing values. The update equation shows that the update process also utilises the existing popularity value for modification. The initial value of $P_f$ for an object is always zero. If the number of queries received for the time period is nix, the popularity of the object is not altered. Otherwise, the popularity of the file increases with the number of queries being received. For every $\delta$ period, the system identifies the possible candidates for replication. This is done by comparing the popularity of a file $P_f(\delta)$ with a threshold popularity value $(P_{th})$. When the popularity of a file at $\delta$ becomes greater than or equal to the threshold value, i.e. $(P_f(\delta) \geq P_{th})$, the process of selecting the target nodes for hosting the replica is initiated.

## B. Site selection and replication

The neighbours and other peers, which are located within 'n' hops, are the candidates for receiving a copy of the replica. The details of the peers of each node are kept in a table called Q-table, which is independently maintained by each node in the network. The Q-table contains the peer addresses along with corresponding ranks being obtained by replicating various files. The ranks are represented as Q-values which depict the performance of the nodes in terms of past replication activities. It is advantageous that objects are shared by a sufficient number of good peers.

*Q-table creation and initialization:* The target nodes are selected from the Q-table. The members for the Q-table are assigned after a simple operation: a message (Hello message) is sent to nodes that come within a time-to-live (TTL) limit, which is the number of hops the message should be propagated; the responded nodes become members of Q-table with some initial Q-value. The message forwarding follows a k-random walk [3] procedure. Initially K messages are generated and the messages are propagated through K number of neighbours selected randomly. Neighbouring nodes forward the message to one of their neighbours; from there to next hop. The message has a message-id. Nodes, which have already received a copy of the message, keep the message-id and address of the neighbouring node to which the message was forwarded. Hence, when a node receives the same message another time it will not be forwarded to a node that has received the message previously, but selects a different peer from the neighbour list. The response messages from the peers consist of equivalent values for their current bandwidth $(b_w)$, and available storage $(s_{avbl})$. Using these values, Q-tables are initialised. The P2P system assigns minimum values for node attributes such as bandwidth $(b_{min})$ and storage $(s_{min})$, which are used for Q-value computation. The Q-values of each node in the table is initialised as $Q_r = \left( \frac{b_w}{b_{min}} + \frac{s_{avbl}}{s_{min}} \right) * 100$. In order to eliminate the random or probabilistic assignment, Q-values are thus initialised with important node attributes such as bandwidth and storage.

*Selection of nodes and replication:* A good peer, which can host a replica, should have a high-speed connection, minimum available storage, link with more number of nodes and it should stay online for a long period. From the possible set of host candidates listed in the Q-table, the best ones according to the bandwidth, available storage, and number of links (degree) are chosen. The objects are copied into nodes, which do not already host the same replica of the target file. Hence, the overwriting of the same file in a node is avoided and at the same time, the process saves bandwidth consumption due to redundant file transfer. In order to choose the possible candidates for hosting the replica, the mean of Q-values listed in the Q-table is computed. Nodes with Q-values greater than or equal to the mean (AvgQ) are selected and a message is sent to each selected node to verify whether a copy of the object exists in their shared folder. *Replication List* of a node is a table that contains a list of object names reserved by other nodes during the object checking process. This evades other nodes to replicate the same object to a node as the same node may be chosen by another node as a target node for replication. If the node is not up, a copy of the object is present, or the object's name appears in the replication list, the node is left out from replicating the chosen file. All other nodes, whose Q-values greater than or equal to AvgQ are selected as target nodes for hosting replicas.

## C. Reward computation

The nodes, which received a copy of the file, send the values for degree $(d_d)$ - the number of neighbours of a node, bandwidth $(b_w)$ and available storage $(S_{avbl})$, after storing the replica to the node that initiated the replication process. This is the reinforcement signal to the replication system. Based on the reinforcement signal, the reward $(\rho_i)$ is computed for each node in the Q-table as

$$\rho_i = \left[ \left( \left[ \frac{d_d}{d_{min} * w_1} \right] + \left[ \frac{b_w}{b_{min} * w_2} \right] + \left[ \frac{s_{avbl}}{s_{min} * w_3} \right] \right) * 100 \right],$$

where $w_1 + w_2 + w_3 = 1$. As the bandwidth is a very important network resource, priority is given for it while computing the reward, hence $w_2 < w_1, w_3$. Therefore, the nodes with large bandwidth highly influence the reward. Moreover sufficient storage space should be available in a node for hosting more and more replicas of different objects. In a P2P network, a few nodes have a large number of degrees while most of other nodes have only a small number of degrees. Peers with a large number of degrees make many replicas as peers with a small number of degrees [16]. In addition, replicas on large degree peers are used frequently as those on peers with small degrees. In our strategy, the system assigns a common minimum degree threshold $(d_{min})$ value to be used for replication to all nodes. In terms of *degree,* the contribution of high degree nodes to the reward is high as compared to low degree nodes. At the same time, nodes with only high bandwidth and storage can also participate in the replication process. All these factors ensure the availability of objects within short hop distances.

## D. Q-table update

The reward values are utilised to modify the Q-values [18]. The update process increase, or decrease the Q-values of peers that are being participated in the replication process. The nodes, which have not participated, do not modify the present Q-values. The nodes with high Q-values are treated as good peers. The Q-values of nodes, which have created a replica, are updated as $Q_{i,t+1} \leftarrow Q_{i,t} + \alpha(\rho_i - Q_{i,t})$, where $\alpha$ is the learning rate (value of $\alpha$ between zero and one), and $Q_{i,t}$ is the present Q-value. If the reward of replication is high, the Q-value is incremented and it relies on bandwidth, available storage and degree of the node. The current Q-values are retained for the nodes comprising a copy of the object i.e. $Q_{i,t+1} \leftarrow Q_{i,t}$. Nodes that are not up are punished heavily with zero reward, $\rho_i = 0$ and the Q-values are updated as $Q_{i,t+1} \leftarrow Q_{i,t}(1-\alpha)$. The assigning high value to the learning rate constant yields a large increase in Q-values of nodes that have placed a replica to their respective directories.

*E. Object replacement*

Some replicas should be deleted to make space for new replicas if adequate storage space is not available in a node. Our replication scheme removes the objects according to their popularity and age. The age attribute represents the time at which the object was inserted into the directory. If the object is recently added to the shared directory, it may have low popularity value and small value for age. Hence, objects with low popularity values and large values for age are removed for housing new objects.

*F. Experimental results*

The performance of Q-replication algorithm is simulated using random graphs that have 10000 nodes. The replication relies on the popularity of the objects. In the simulation scenario, all the queries contain keywords alone. Each node generates 100 queries and one query is propagated every 20 seconds on average. However, each node enters the query generation phase in a randomly selected time slot. Hence, the flood of query message production is regulated. Eighty percent of the nodes are up at the time of performing simulation. Fifty percent of 'Down' nodes selected randomly change their status to 'UP' after every 50,000 queries are propagated and, at the same time, the same amounts of UP nodes obtain the DOWN status. The default TTL value is preset as six. The effect of Q-replication on Distributed Search Technique (DST) [17] and random walk [3] is evaluated.

*Availability of objects:* Initially, there are hundred objects distributed to various nodes. In order to study the effect of Q-replication on availability of objects, simulations are conducted with the condition that the objects, which are discovered during searching (DST), are not copied into the requester node, however the popularity and hit rate are modified as the result of searching. This is employed for counting the number of replicas being created by the Q-replication scheme. The popularity threshold $P_{th}$ is preset as five for simulation, i.e. objects with popularity greater than or equal to five are chosen for replication. The number of queries propagated through the network increases, the object availability, also increase (figure 1). For less number of queries, there are less number of objects present on nodes. As the availability of objects relies on the popularity of the objects, queries that are more successful increase the quantity of objects in each active node.

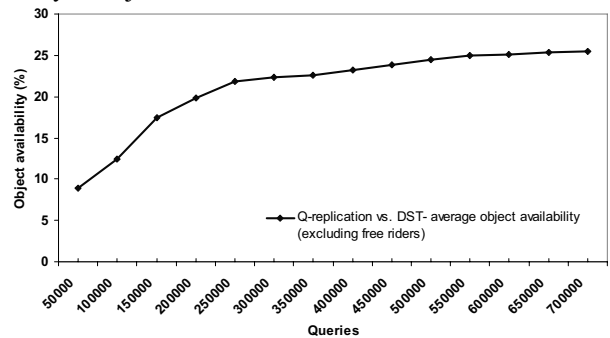

**Figure 1: Object availability**

*Churn rate:* The Q-replication surmount the high rate of node dynamics. Churn arises from continued and rapid arrival and failure (or departure) of a large number of peers in the P2P system. It has the potential to increase host loads and block a large fraction of normal search operations in the system. The Q-replication ensures high resistance to churn attacks and this is evident from figure 2 in which in even high rate of churn, the system generates a good success rate. This is made possible by replicating the popular objects to good number of peers for mounting the availability. The graph shows different situations where N% of nodes are down.

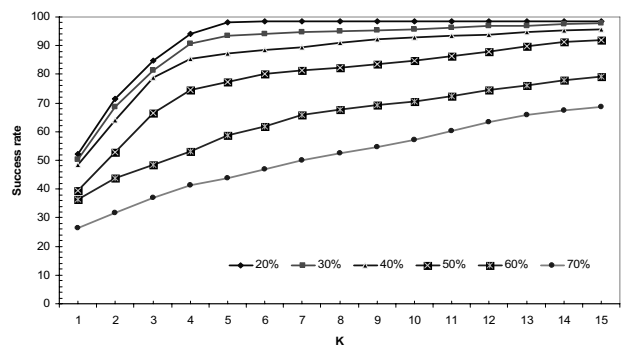

**Figure 2: Success rate and churn rate (DST)**

*Queries finished (random walk):* Simulations are conducted in a random network comprising 10000 nodes to compare the performance of path replication [3] and Q-replication using random K-walk. The number of walkers are limited to six. The results for queries finished for varying TTLs are shown in figure 3. The performance of Q-replication is superior to path replication in each interval and it relies on the TTL value being used. Q-replication creates replicas of objects in more well-performing nodes and at the same time, path replication relies only the nodes on the search path in which the nodes are not selected based on their performance.

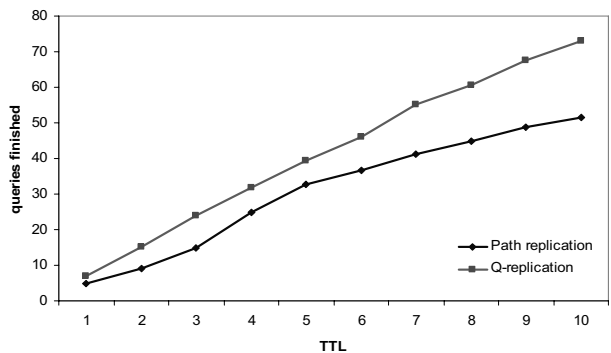

**Figure 3: Percentage of queries finished**

## IV. Conclusions

In this paper, a few existing replication techniques are discussed. A Q-learning based replication scheme for proactively deploying data replicas on several other peers is presented. The aim of the technique is to decrease the consumed messages before hit and improvement of the hit rate. The approach utilises the popularity of the objects and the objects are distributed with more copies to various sites based on the site selection logic. The popularity is computed according to the queries received on a particular objects and the total number of queries received by the node for a certain period. The target nodes are selected not randomly nor probabilistically, but they are chosen based on their past performance. The replication scheme does not rely on nodes on the search path. Other nodes can also host the same replica of the object, provided that the sites satisfy certain criteria. The replacement of a file follows a different approach and it depends on its popularity and age. The reward calculation and the Q-value update processes are explained. The Q-replication follows a popularity based replication policy. Hence, the availability of popular objects increases with time. The Q-replication performance is also compared with path replication using random walk algorithm in a random network. The important property of replication algorithm, coping with churn rate is evaluated with different churn rate. The Q-replication assists the distributed search scheme to produce more successes during large churn rate.